\documentclass[aps,prl,showpacs,reprint,superscriptaddress,twocolumn]{revtex4-2}
\usepackage[english]{babel}
\newcommand{\beginsupplement}{%
        \setcounter{table}{0}
        \renewcommand{\thetable}{S\arabic{table}}%
        \setcounter{figure}{0}
        \renewcommand{\thefigure}{S\arabic{figure}}%
     }
\usepackage{graphicx}
\usepackage{color}
\usepackage{amsmath}
\usepackage{bbm}
\usepackage{amssymb}
 \usepackage{array}

\usepackage{dsfont}

\usepackage[utf8]{inputenc}	
\usepackage[T1]{fontenc}

\usepackage[svgnames]{xcolor}

\input epsf

\begin{document}

\title{Signatures of superconducting pairing driven by electron-electron interactions in moiré WSe$_2$/WSe$_2$ homobilayer modelled by Hubbard Hamiltonian}
\author{Andrzej Biborski}
\email{andrzej.biborski@agh.edu.pl}
\affiliation{Academic Centre for Materials and Nanotechnology, AGH University of Krakow, Al. Mickiewicza 30, 30-059 Krakow,
Poland}
\author{Micha{\l} Zegrodnik}
\email{michal.zegrodnik@agh.edu.pl}
\affiliation{Academic Centre for Materials and Nanotechnology, AGH University of Krakow, Al. Mickiewicza 30, 30-059 Krakow,
Poland}

\begin{abstract}
Strong evidence of unconventional superconductivity has been very recently reported experimentally in twisted transition metal dichalcogenide bilayer and gathered a significant amount of interest. Here we consider the Hubbard model on a triangular lattice describing the hole-doped moiré superlattice emerging in WSe$_{2}$/WSe$_{2}$ twisted homobilayer in the moderately correlated regime. By applying the Density Matrix Renormalization Group, we diagonalize the spin-valley-polarized Hamiltonian and show signatures of coexisting singlet and triplet pairings in the range of hole dopings and displacement fields reported in the experiments. In this view, we show that the superconductivity in the WSe$_{2}$/WSe$_{2}$ twisted homobilayer is likely to be induced by electronic correlations and has a mixed-symmetry character. These predictions can shed light on the nature of the superconducting state observed in the twisted homobilayer of WSe$_{2}$/WSe$_{2}$. We also identify the emerging superconducting orders, which are $d_{xy}(d_{x^2-y^2} \pm id_{xy} )$ and $p_y(p_{x}\mp ip_{y})$ for the singlet and triplet channels in the  cylinder of width three(four), respectively.
\end{abstract}

\maketitle

\emph{Introduction.} Superconductivity in strongly correlated electronic systems is still a mysterious phenomenon and one of the main unresolved problems in condensed matter physics~\cite{Cao2018_2,Spalek2022,Stewart03042017}. Nevertheless, the two-dimensional character of pairing emerging in the presence of strong electron-electron interactions compared to the kinetic component is believed to be a common feature of unconventional superconductors. Therefore, it is tempting to examine the possible pairing and its nature in \emph{in-situ} controllable systems as an alternative to \emph{traditional} systems such as cuprates~\cite{Proust2019,Keimer2015}. 
Such a simulation platform can be achieved by artificially coupling two or more layers of two-dimensional material mismatched by the relative twist (homobilayers) or structure/chemical composition (heterobilayers). This results in the formation of so-called moiré superlattices in which the potential confining itinerant carriers is modulated with a length of an order of magnitude greater than the lattice constant of a single layer. These systems have recently been investigated with both experimental and theoretical approaches as provided in a number of papers~\cite{Chen2019,Balents.2020,Mak.2022,Kennes.2021,Cao2018_1,Cao2018_2,Rademaker_2023_flat_bands,Shen2020,Wang2020,Ghiotto2021,Wu2018,Haining2022,Xu2022,Sheng2023,Millis2023,Zegrodnik2023,Guo2025}, revealing the emergence of a variety of exotic states believed to be driven by electronic correlations. 

In particular, systems composed of transition metal dichalcogenides (TMDs) are of great interest because of their tunability and robustness of narrow bands compared to twisted bilayer graphene~\cite{Haining2020}. That is, the metal-insulator transition (MIT)~~\cite{Ghiotto2021} and the emergence of the superconducting state (SC) have been reported experimentally in  WSe$_2$/WSe$_2$ homobilayer~~\cite{Xia2025,Guo2025}. Furthermore, generalized Wigner crystals (GWCs) have been found in various fractional fillings in the WSe$_2$/WS$_2$ heterobilayer~~\cite{Li2021} indicating the importance of electron-electron correlations in this class of materials. Although the presence of MIT and GWC can be understood along the paradigms of strongly correlated physics, the nature of SC in WSe$_2$/WSe$_2$ is not fully recognized since it appears in samples where the opening of the Mott gap is not detected in the regime where SC appears~~\cite{Xia2025,Guo2025}. The presence of spin-orbit polarization and topological features encoded in this system additionally makes the analysis more complex, as we provided theoretically recently~\cite{Zegrodnik2023,Zegrodnik2024} based on the variants of $t-J$ model.

In this Letter, we state the question whether the SC state in WSe$_2$/WSe$_2$ homobilayer can emerge in the picture of \emph{archetypical} Hubbard model which is supplied with a non-interacting part of Hamiltonian, which effectively models spin-valley polarization of the narrow band. As is commonly known, the Hubbard model formulated for the two-dimensional lattices delivers astonishingly diverse phase diagrams despite its simplicity~~\cite{Arovas2022,Qin2022}. The single band Hubbard model is proven to describe consistently the MIT in strongly correlated systems. Nevertheless, it is still debated whether it can capture the nature of superconductivity in cuprates when only a single band of the underlying square lattice is taken into account~\cite{Qiu2020,Jiang2023}. However, the minimal model  of the WSe$_2$/WSe$_2$ homobilayer is defined on the triangular lattice~\cite{Haining2020,Wang2020}, the band is spin-split and the magnitude of Hubbard on-site interaction $U$ is likely to be of order of the width of the narrow conduction band; thus,the lack of SC pairing pointed out by Qin et al.~~\cite{Qiu2020} does not necessarily apply to this context. 

\begin{figure}
    \centering
    \includegraphics[width=0.95\linewidth]{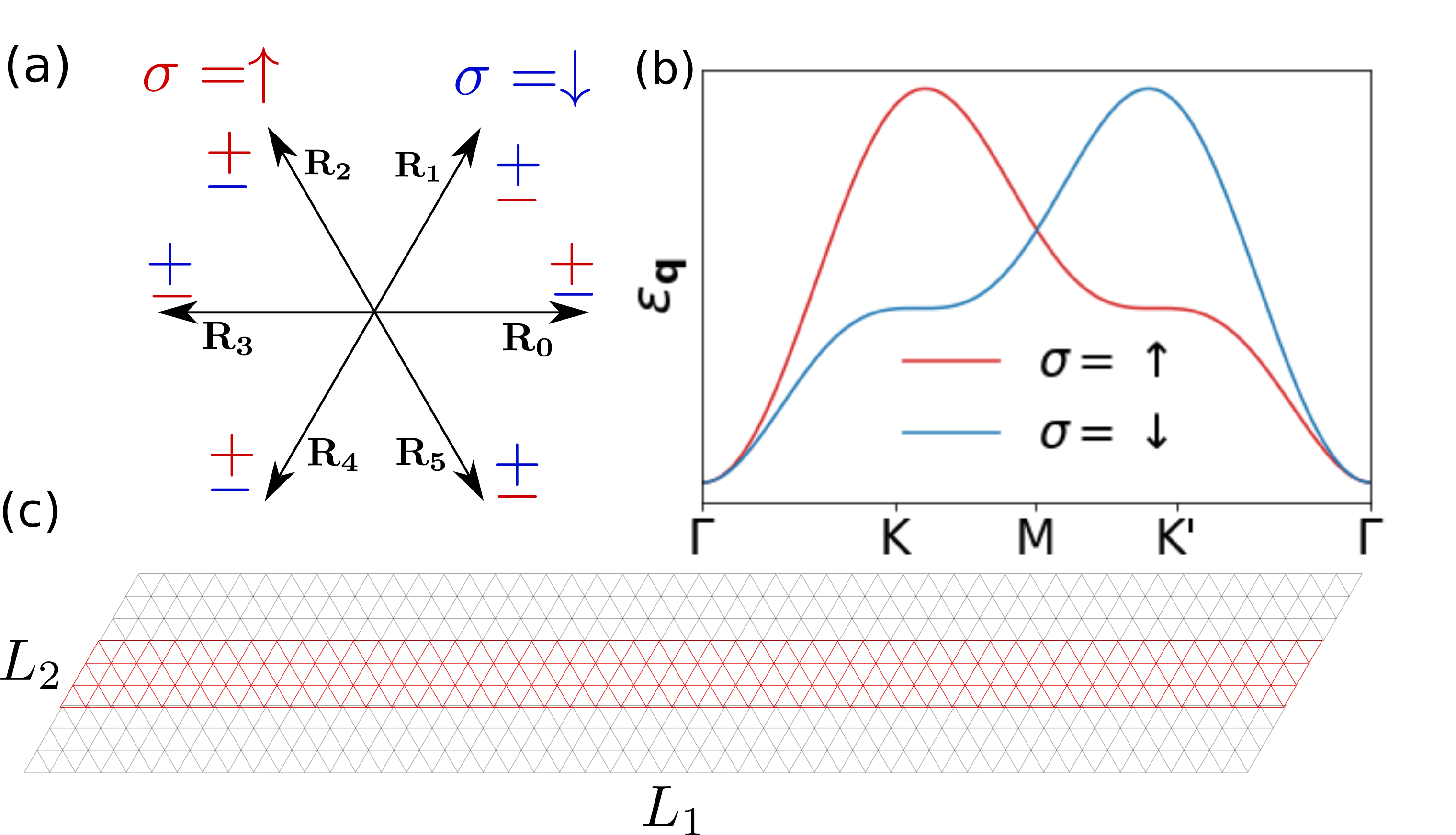}
    \caption{ (a) The sign of phase hopping with respect to the direction $\mathbf{R}_i$ and spin $\sigma=\{\uparrow,\downarrow\}$; (b) Spin-split bare-band for hopping at $D=0.4$ V/nm; (c) The cylinder of size $L_1\times L_2$ (red area) considered in DMRG calculations, shaded parts  are replicas of simulated sample and symbol periodic boundary conditions applied along $\mathbf{R}_1$ direction.}
    \label{fig:hopping_scheme}
\end{figure}

\emph{Model and method.} Stimulated by recently reported superconductivity in WSe$_2$/WSe$_2$ bilayer system ~~\cite{Xia2025,Guo2025} we provide indicators of the presence of mixed singlet-triplet pairings that are obtained in the framework of the density matrix renormalization group method (DMRG) in its matrix product state (MPS) formulation. 
Therefore, our findings suggest that the SC state in WSe$_2$/WSe$_2$ homobilayer can be driven by electron-electron correlations when the Hubbard model paradigm is utilized. Namely, we consider a Hubbard type Hamiltonian on a triangular lattice defined as
\begin{align}
    \mathcal{\hat{H}}=\sum_{\langle i,j\rangle}\sum_{\sigma}t_{ij}^{D,\sigma}\hat{a}^{\dagger}_{i,\sigma}\hat{a}_{j,\sigma}+\sum_{i}U\hat{n}_{i\uparrow}\hat{n}_{i\downarrow},
    \label{eq:H}
\end{align}
where the hopping amplitude $t_{ij}^{D,\sigma}$ is a complex number whose absolute value depends on the displacement field $D$ applied  perpendicularly to the sample in the experiment. The phase of hopping depends on $D$, spin $\sigma$, and the direction of the vector pointing from the site $i$ to one of its six neighboring sites $j$. The phase sketch is depicted in Fig.\ref{fig:hopping_scheme}a and the dependency of the hoppings on $D$ is taken from Ref.~\cite{Wang2020} and corresponds to the twist angle $\theta=5.08^{\circ}$. The resulting bare band structure is eventually characterized by spin-valley polarization, as shown in Fig.\ref{fig:hopping_scheme}b.\begin{table}[h]
    \centering
    \begin{tabular}{|c || c | c | c | c | c | c | c|}
        \hline
        $D$ (V/nm) & 0.25 & 0.30 & 0.35 & 0.40 & 0.45 & 0.50 & 0.55 \\
        \hline
        $-\text{Re }t^{D,\uparrow}_{i,j(\mathbf{R}_i+\mathbf{R}_0)}$ (meV) & 7.5 & 7.4 & 7.3 & 7.1 & 6.9 & 6.8 & 6.7 \\  \hline
        $\text{Im }t^{D,\uparrow}_{i,j(\mathbf{R}_i+\mathbf{R}_0)}$ (meV) & 3.2 & 3.7 & 4.3 & 4.7 & 5.3 & 5.8 & 6.3 \\
        
        \hline
    \end{tabular}
    \caption{Values of displacement field and related hopping amplitudes utilized here as provided in Ref.~\cite{Wang2020}.}
    \label{tab:hoppings}
\end{table}
$U$ is standard Hubbard repulsion, which can be related to the energy scale given by bare-bandwidth $W\approx9|t_{ij}^{D,\sigma}|$. In experiments reporting the superconducting state in the WSe$_2$/WSe$_2$ system, the MIT has not been detected in the entire phase diagram~~\cite{Guo2025} or in the regime where the superconductivity appears~~\cite{Xia2025}. At the same time, clear signatures of MIT have been observed in other reports~\cite{Ghiotto2021,Wang2020}. Therefore, we present data obtained mainly for $U=0.08$ eV, which is slightly below $W$ (cf. the hopping values given in Tab.\ref{tab:hoppings}). However, our selection of $U$ is close to the value for which MIT is supposed to appear in the Hubbard model on the triangular lattice but supplied with the spin-valley polarized single-particle terms~\cite{biborski2024}. In this respect, our choice of $U$ appears to be close to all the scenarios reported experimentally, as we identify it to be on the boundary between moderately and strongly correlated regimes. We employ the DMRG method to find an approximate ground state of the Hamiltonian given in Eq.~\ref{eq:H} mainly for the $L_1\times L_2=48\times 3$ cylinder, applying open (in $\mathbf{R_0}=(1,0)$ direction) - periodic (in $\mathbf{R_{1}}=(\frac{1}{2},\frac{\sqrt{3}}{2})$ direction) boundary conditions (Fig.~\ref{fig:hopping_scheme}c). 
The hoppings are complex numbers, which results in an increased size of the MPS and a Hamiltonian tensorial representation. Therefore, we take a relatively small value of $L_2=3$ and perform the analysis in a direction parallel to $\mathbf{R_0}$. However, we have also  performed the auxiliary calculations for  the cylinder width $L_2=4$ for the selected parameters to validate the obtained results in view of the narrow cylinder shape assumed. The majority of presented data is obtained for the maximal bond dimension $M=12000$ using the TenPy library~\cite{tenpy2024,biborski_andrzej_zenodo}, which resulted in a maximal truncation error of the order $10^{-6}$. In other cases concerning $L_2=4$, the truncation error is of the order of $10^{-5}$ for the maximal value of $M=14000$.

\emph{Results.} To explore the superconducting  properties we first focus on the analysis of pair-pair correlation functions ~\cite{Yang1962,Anderson1966,Tsuei2000,Daul2000,Sheng2023,Chen2024} $P_{\nu}^{\alpha\beta}(n)\equiv\langle\hat{\Delta}^{\dagger}_{\nu,\alpha}(\mathbf{R}_i)\hat{\Delta}_{\nu,\beta}(\mathbf{R}_i+n\mathbf{R}_0)\rangle$, where $\hat{\Delta}_{\nu,\alpha}(\mathbf{R}_i)=\hat{a}_{\mathbf{R}_{i},\uparrow}\hat{a}_{\mathbf{R}_{i}+\mathbf{R}_\alpha,
\downarrow}-(-1)^{\nu}\hat{a}_{\mathbf{R}_{i},\downarrow}\hat{a}_{\mathbf{R}_{i}+\mathbf{R}_\alpha,\uparrow}$, and $\nu$ indicate singlet ($\nu=0$) or triplet ($\nu=1$) channel in pairings. Note that $\alpha$ ($\beta$) numbers vector $\mathbf{R}_{\alpha}$ ($\mathbf{R}_{\beta}$).  We focus on the hole-doping $\delta=N_{el}/L_1 \times L_2$ - where $N_{el}$ stands for the number of electrons - $\in(~0.9,1.0)$ and displacement field $D\in(2.5,5.5)$ V/nm - approximate values for which superconductivity in WSe$_2$/WSe$_2$ has been reported~\cite{Guo2025}. 
\begin{figure}
    \centering
    \includegraphics[width=0.9\linewidth]{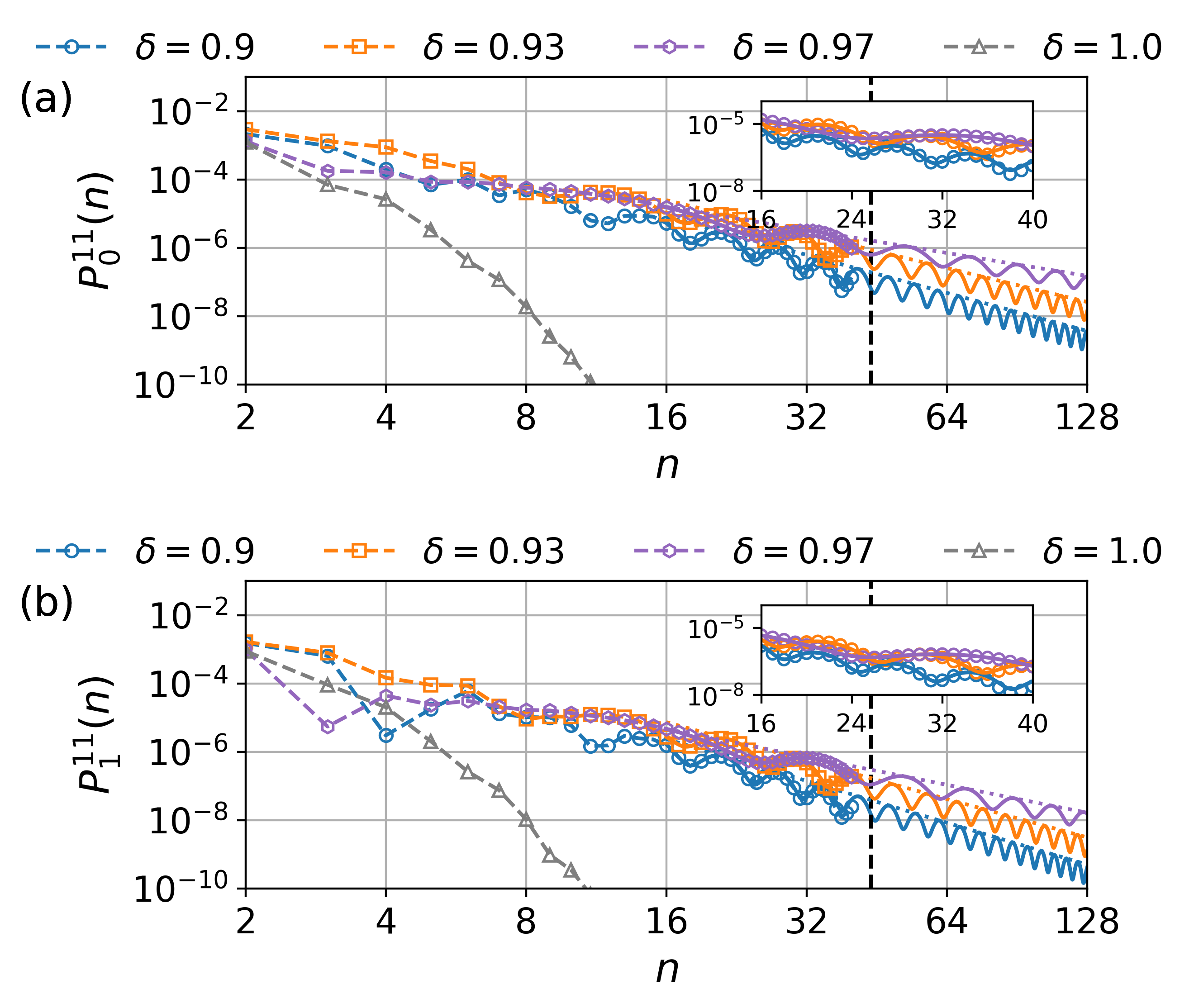}
     \caption{ The decays of singlet (a) and triplet (b) pairing correlation functions at $D=0.4$V/nm obtained for the hole doping in the vicinity of half-filled band. Symbols refer to the obtained data from DMRG. The solid lines are maxima of fits obtained for $n\in[16,40)$ and further extrapolated beyond $L_1$ (marked as vertical dashed lines). In the insets the zoom of range  for which  the fitting procedure have been carried out is shown indicating high quality of power-law decaying fits.}
    \label{fig:4.0-decays}
\end{figure}

\begin{figure}
    \centering
    \includegraphics[width=0.95\linewidth]{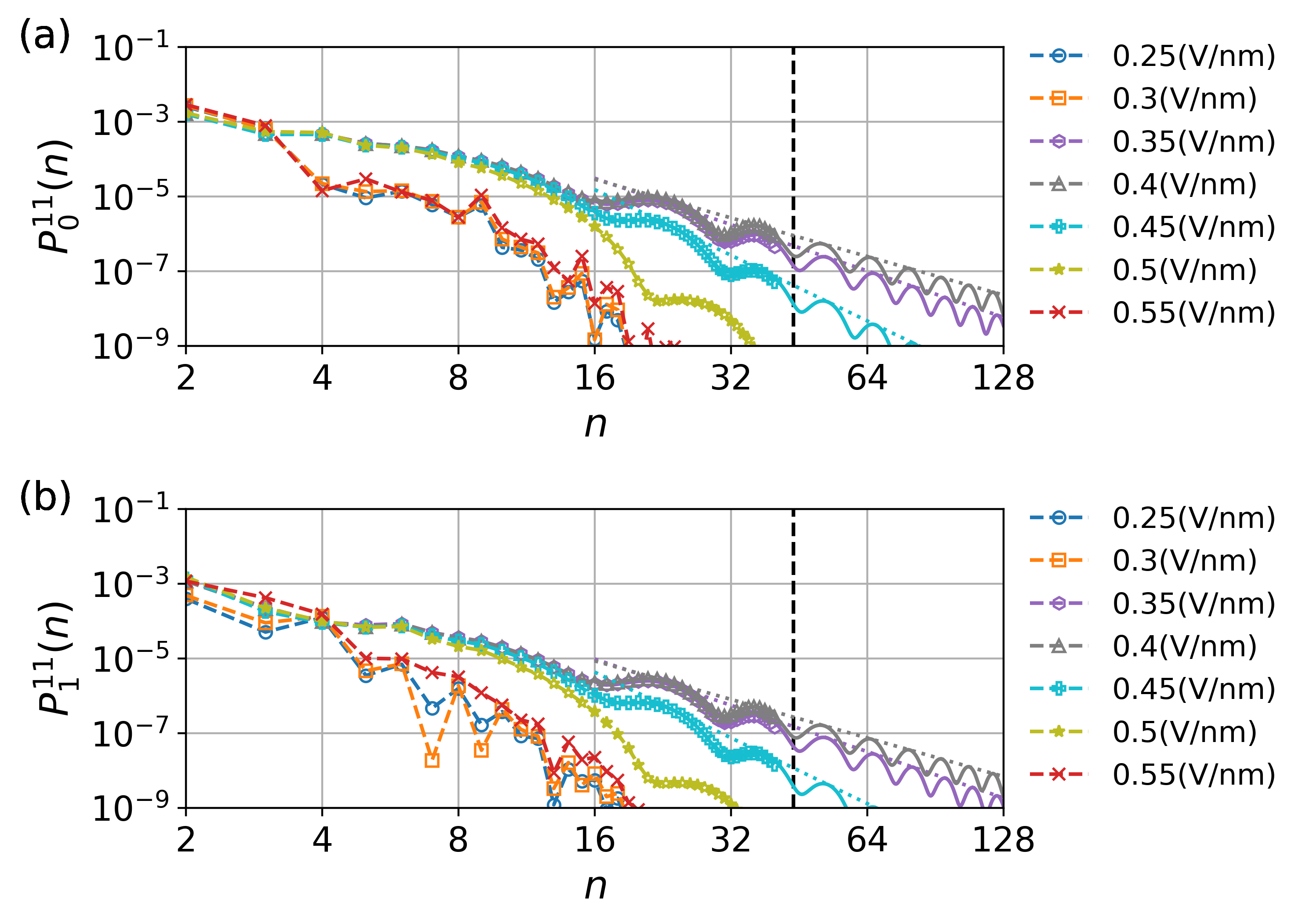}

    \caption{ The decays of singlet (a) and triplet (b) pairing correlaton functions at $\delta\approx0.96$ collected for $D\in[0.25,0.55]$ V/nm. The lines and symbols meaning is the same as in Fig.\ref{fig:4.0-decays}.}
    \label{fig:D-decays}
\end{figure}

In Fig.\ref{fig:4.0-decays} we present the decay of pairing correlation functions collected at $D=0.4$ V/nm. To reduce the bias effect of open boundary conditions in the calculation of $P_{\nu}^{\alpha\beta}$, we take the reference site $\mathbf{R}_{i}=4\mathbf{R}_0$ and perform the analysis of the maximal distance referring to $n=40$~\cite{supp}. The decay in $P_{\nu}^{11}$ (excluding the $\delta=0$ case) is of oscillatory character, since periodic charge modulation emerges in the fillings considered~\cite{White2002}. We observe a clear difference among the values of correlation functions with respect to $\delta$ when $n$ increases. For $\delta=0.92-0.97$, the magnitude in pairings is clearly stronger than for higher values of hole doping. In addition, we observe a rapid decrease in $P_{\nu}^{11}(n)$ for $\delta>0.97$; that is, for $n\gtrsim16$, it becomes less than $10^{-10}$. Importantly, both singlet and triplet channels participate in pairings for $0.92\lesssim\delta\lesssim0.97$. The role of $D$ is also prominent in view of the formation of a quasi-long-range superconducting order, as can be deduced from Fig.\ref{fig:D-decays}.
That is, at $\delta\approx0.96$ we obtain a clearly enhanced signal in $P_{\nu}^{11}$ for $0.35$ V/nm $\lesssim D\lesssim 0.45$ V/nm in both singlet and triplet channels compared to $D\lesssim0.35$ V/nm and $D\gtrsim4.5$ V/nm. 

\begin{figure}
    \centering
    \includegraphics[width=1.1\linewidth]{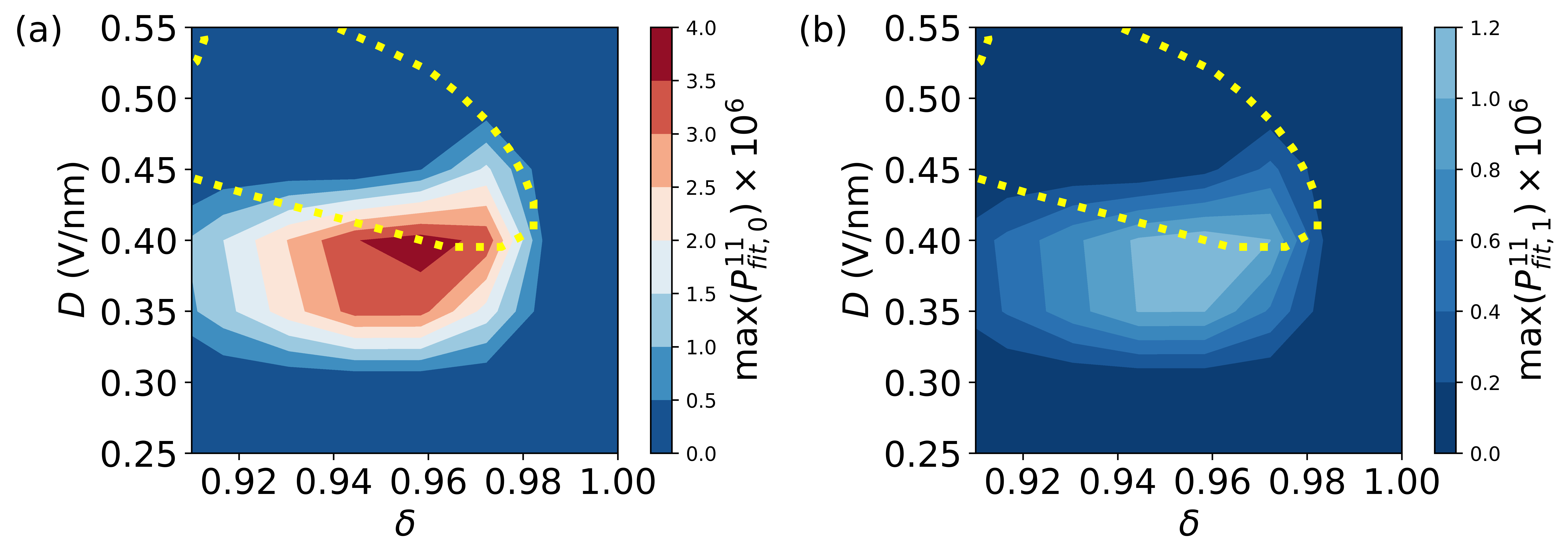}
    \caption{The interpolated maps representing  phase diagram of pairing. That is  maximal value of $P^{1,1}_{fit,\nu}(r=32|\mathbf{R_0}|)$ for the selected ranges of doping and displacement fields for singlet (a) and triplet (b) channels. The yellow dotted lines denote region for which superconductivity has been detected in experiment performed by Guo et al.~\cite{Guo2025}.   }
    \label{fig:domesmap}
\end{figure}

As we intended to characterize the tendency towards superconducting order in a more quantitative manner, we fitted the curves of the form ${P}_{fit,\nu}^{\alpha\beta}(r)=\big[A\cos{(Br+C)}+D\big] \kappa(r)$ where $\kappa(r)=r^{-K_s}$ where $K_s$ is the Luttinger exponent or, alternatively, $\kappa(r)=\text{e}^{-r/\xi_s}$ \cite{Sheng2023}. This allowed us to at least partially exclude the biasing effect of oscillations with respect to the selection of the position of the reference site, as well as to identify the emergence of quasi long-ranged superconducting order. To achieve this, in Fig.~\ref{fig:domesmap} we compare  \emph{upper-bound} values of ${P}_{fit,\nu}^{\alpha\beta}(r)$  at $r=32|\mathbf{R}_0|$.

\begin{figure}
    \centering
    \includegraphics[width=0.8\linewidth]{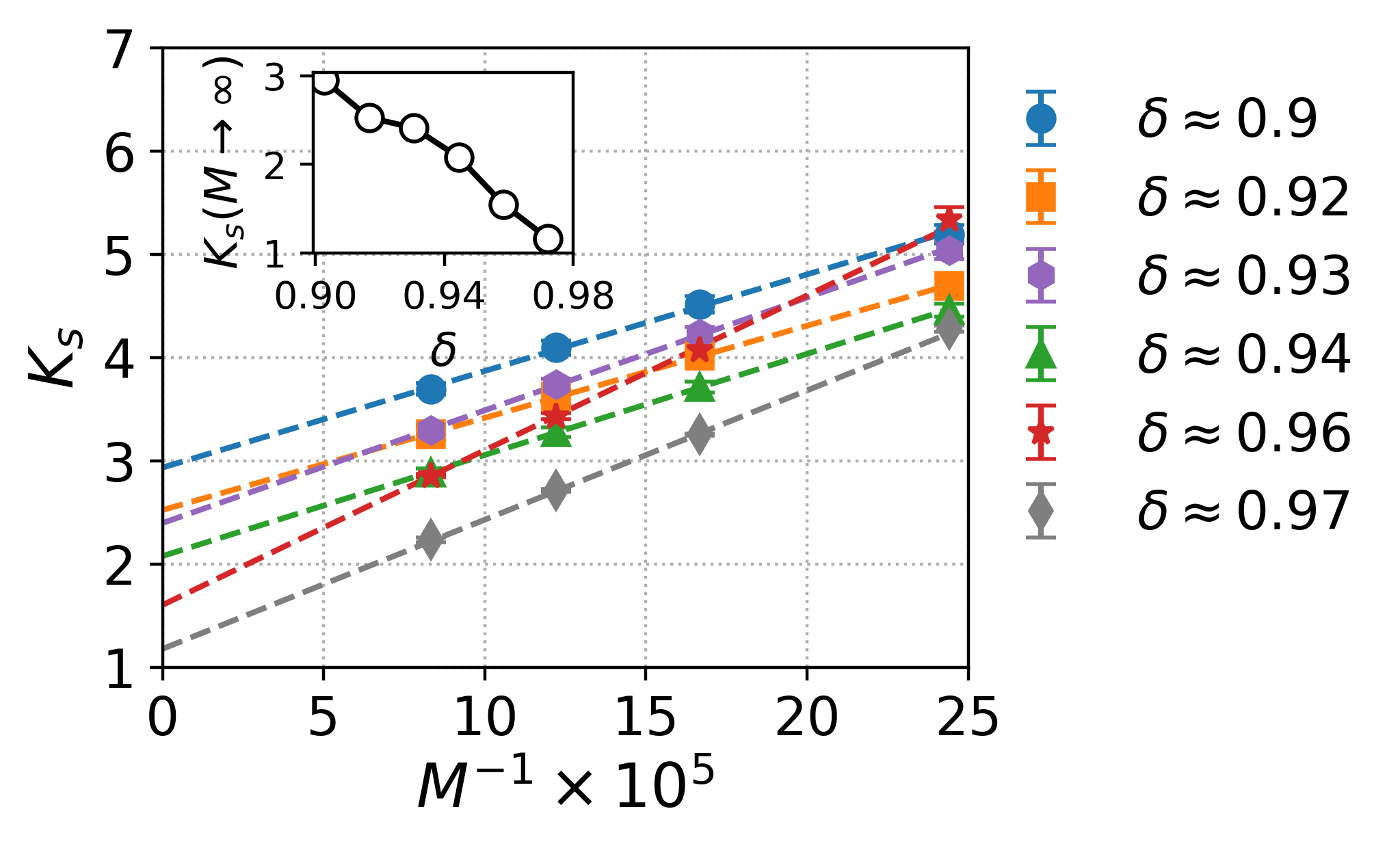}
    \caption{The power-law decay exponent $K_s$ at $D=0.4$ V/nm for the considered dopings as of function of $1/M$. The lines are linear fits and in the inset virtually exact exponents, that is $K_s(M\rightarrow\infty)$ are shown. Presented data refer to the singlet pairing since for the triplet channel we obtain nearly identical outcome, note, that estimated errors are smaller than symbol size.  }
    \label{fig:ks}
\end{figure}

\begin{figure}
    \centering
    \includegraphics[width=0.8\linewidth]{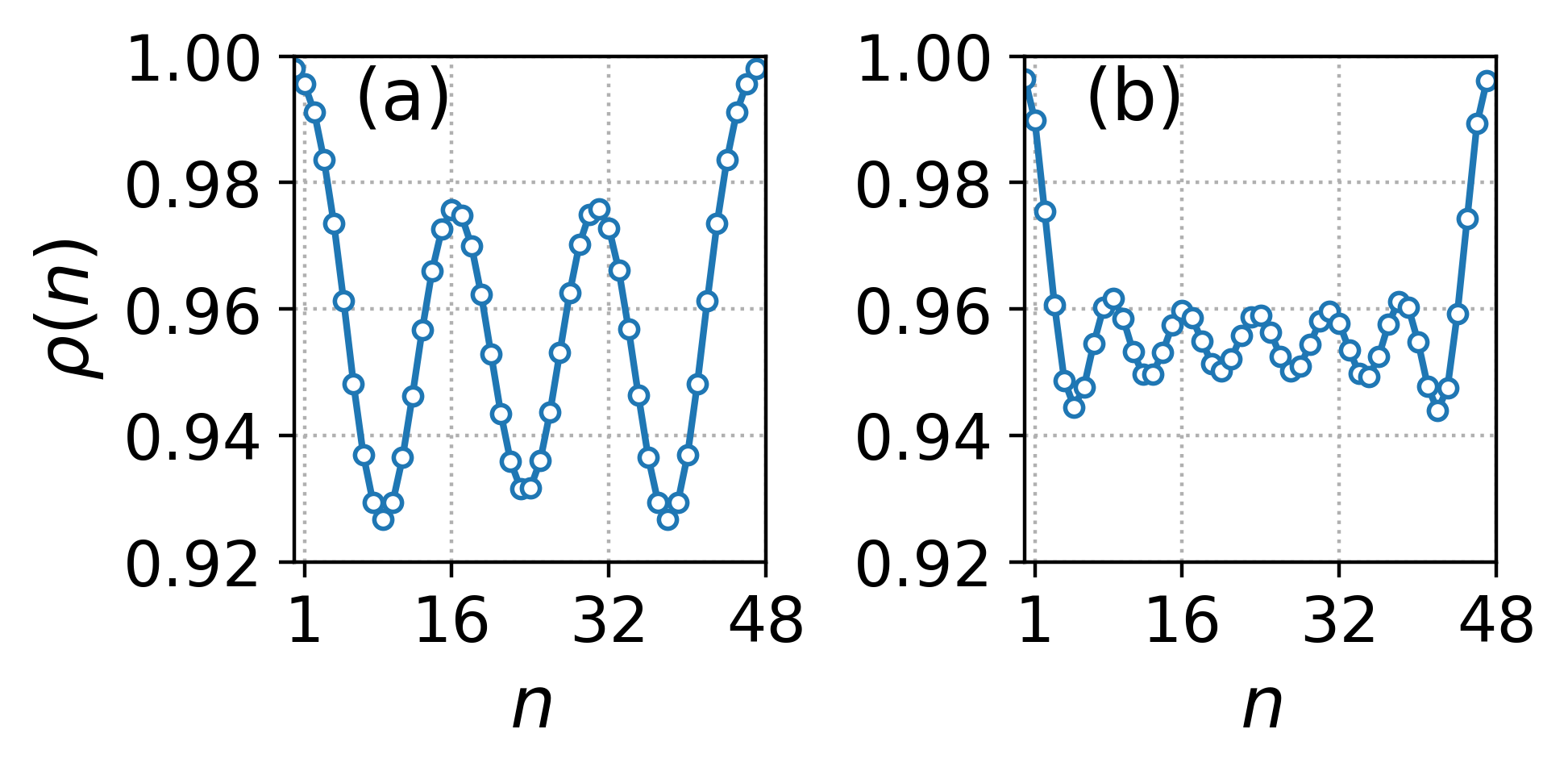}
    \caption{Density profiles obtained for  $D=0.4$ V/nm (a) and $D=0.25$ V/nm (b) at doping $\delta\approx0.96$. For $D=0.4$ V/nm the period of oscillations is the same as for pairing correlation functions.  }
    \label{fig:densities}
\end{figure}

We find the power-law decay of correlation functions when $0.35$ V/nm $\lesssim D\lesssim 0.45$ V/nm and $\delta\lesssim  0.98$, that is, where the pairing signal is highest. In the remaining situations, the decay is of exponential character. The singlet and triplet pairings coexist in the same range of $\delta$ and $D$; however, the latter pairing is characterized by an amplitude that is noticeably smaller than the former. The area of the phase diagram with enhanced pairing amplitudes is astonishingly close to the superconductivity reported in the experiment carried out by Guo et al.~\cite{Guo2025} as presented in Fig.\ref{fig:domesmap}. In particular, we observe the SC phase only in a specific range of non-zero displacement fields and an abrupt suppression of superconductivity at half-filling. Note that this result is in contrast to the other experimental report by Xia et al.~\cite{Xia2025}, which, however, corresponds to a significantly different twist angle $\theta\approx3.5^{\circ}$ than considered both by us here and in Ref.~\onlinecite{Guo2025} ($\theta\approx 5^{\circ}$). It should be emphasized that the magnitude of the pairing correlation functions increases with increasing number of maximal bond dimensions, especially for larger $n$. Namely $K_s\propto1/M$, as can be deduced from Fig.\ref{fig:ks}. The minimal value of $K_s\approx1$, which is attributed to the strongest enhancement in pairing, results from the extrapolation $M\rightarrow\infty$ for $D=0.4$ and $\delta\approx0.97$ (see the inset in Fig.\ref{fig:ks}). 
The wavelength of oscillations in the pairing correlation function is $\lambda\approx L_1/3$ when the decay is detected to conform to the power law behavior. This is in accordance with the charge density profile ${\rho}(n)=\big\langle \sum_{\sigma}\hat{a}_{n\sigma}^{\dagger}\hat{a}_{n\sigma}\big\rangle$, as shown in Fig.\ref{fig:densities}a. We also observe clear charge modulations when the decay in pairing is exponential. However, the period of $\rho(n)$ for given $\delta$ changes with respect to $D$, e.g. at $\delta\approx0.96$ and $D=0.25$ V/nm $\lambda\approx L_1/6$ (Fig.\ref{fig:densities}b).  This may suggest that Friedel oscillations~\cite{Bedurftig1998,White2002,Jiang2021} driven by open boundaries along $\mathbf{R}_0$ are responsible for the periodic modulation in $\rho$. The issue of whether the nature of charge order observed is (or is not) the artifact of cylindrical shape should be clarified with further studies concerning the model utilized here.

Up to this point, we have analyzed pairings related to $(\alpha,\beta)=(1,1)$ only. However, inspection of pairs which are not aligned in parallel, that is $\alpha\neq\beta$, allowed us to further characterize the quasi-long-range superconducting order. Namely, the identification of pairing order can be achieved on the basis of the real-space resolved  superconducting order parameter $\phi_{o}=\sqrt{\lim_{n\rightarrow\infty}\phi_{o}^{2}(n)}$,  where
\begin{equation}    
    \phi_{o}^{2}(n)=\Bigg|\sum_{\alpha,\beta}f_{o}^{\alpha}(f_{o}^{\beta})^{*}P_{\nu(o)}^{\alpha,\beta}(n)\Bigg|    
\end{equation}
$f_{o}^{\alpha}$ are \emph{form-factors}  with $o\in\{s,d_{xy},d_{x^2-y^s}\}\cup\{d_{x^2-y^2} \pm id_{xy}\}$, and, $o\in\{p_x,p_y,f\} \cup \{p_y\pm ip_{x}\}$ orders in the singlet and triplet channels, respectively (see Supplemental Material\cite{supp} and Refs. ~\cite{Koretsune2005,Wu_2013,Zegrodnik2023} for the details). Note that $d_{x^2-y^2} \pm id_{xy}\equiv d\pm id$ and $p_x\pm ip_{y}\equiv p\pm ip$ are extensions of the order parameter sets to the complex number domain, rendering them to conform to the $C_3$ symmetry as the considered spin-valley polarized Hamiltonian. We emphasize also that as we work in a regime assuming a constant number of particles for a given $\delta$ we \emph{need} to inspect the square of $\phi_{\nu,o}(n)$ as it is the expectation value of the operator conserving the number of electrons considered. 

First, we have identified the superconducting order by direct inspection of ratios between pairing functions for $n=16$ which revealed the formation of $d_{xy}$ and $p_y$ order as shown in Fig.~\ref{fig:formfactors}. This allowed us also to set the proper \emph{gauge} in the order parameter (see Supplemental Material)  definition. Eventually,
in Figs.~\ref{fig:scorder48}a,b we present $\phi_{\nu,o}^{2}(n)$ for the cylinder of shape $L_1\times L_2=48\times3$ at $\delta\approx0.96$ and $D=0.04$ V/nm. 
The detailed inspection of pairing order parameters provides $\phi_{d+id}\approx \phi_{d-id}\approx\phi_{d_{xy}}^{2}$, therefore the dominating superconducting order is $d_{xy}$ in the singlet channel, since $\phi_{d_{x^2-y^2}}^{2}$ and $\phi_{s}^{2}$ are almost zero for the whole range of considered $n$. Explicitly, the  magnitudes  of $\phi_{d\pm id}$ are dominated by $d_{xy}$ component, which we have checked also by direct analysis of the relations among magnitudes and phases of $P_{\nu}^{\alpha\beta}$ correlation functions as stated above. Also, the ratio $\phi_{d\pm id}^2/\phi_{d_{xy}}^2$ is nearly exactly $3/4$ - that is the same value as that resulting by assuming pure $d_{xy}$ pairing in the singlet channel. This observation again indicates that the magnitude of $\phi_{d\pm id}$ comes solely from $d_{xy}$ contribution. The similar analysis carried out for the triplet channel confirmed the tendency toward the formation of coexisting $p_y$ order, as can be deduced from Fig.~\ref{fig:scorder48}b. 

The symmetries obtained  are slightly different from those we identified in our recent study based on the $t-J-U$ model~\cite{Zegrodnik2023} where we have found coexisting $d\pm id$ and $p\mp ip$ orders. However, here we consider not only a different model and apply a different numerical approach, but also simulate the system with \emph{explicitly broken symmetry} that is encoded in the boundary conditions and the shape of the cylinder effectively described by the MPS, which is formally one-dimensional ~\cite{SCHOLLWOCK201196}. 

According to the latter issue, which is mostly supposed to be responsible for the $C_6$ symmetry breaking in the absolute values of pairings, it is tempting to investigate if an increase in $L_2$ may provide a more isotropic type of order. While the magnitude of correlation functions is prone to bond dimension $M$ and with increasing $L_2$ the minimal value of $M$ for which pairings are detectable increases as well, the problem is computationally challenging. Thus, the true \emph{bulk} properties of the system are hardly obtainable in terms of the standard DMRG-MPS formulation. 

Nevertheless, we have found the $L_1\times L_2=24\times4$ cylinder as achievable with respect to the satisfactory numerical accuracy. Although a shorter cylinder with $L_2=24$ provides less information  if quasi-long-range pairing is robust, that is, that correlations decay algebraically, it  provides an insight into the kind of order developed in the shorter range when $L_2$ is increased -  as we show in Figs.~\ref{fig:scorder24}a,b. Indeed, we observe  the development of $d+id/p-ip$ superconducting pairings, since $\phi_{d+id}^{2}>\phi_{d_{xy}}^{2}, \phi_{d_{x^2-y^2}}^{2}$ and $\phi_{d-id}^{2}\approx0$ in the singlet channel, and $\phi_{p-ip}^{2}>\phi_{p_y}^2,\phi_{p_x}^2$ and $\phi_{p+ip}^2\approx0$ in the triplet channel. Note that these relations hold irrespective of the chosen \emph{gauge} for the order parameter, which, however, does not influence by construction the magnitudes of $\phi_{d\pm id}^2$ and $\phi_{p\pm ip}^2$. These observations point out  the possible role of cylinder width in view of the form of pairing order. Furthermore, they suggest that  in \emph{bulk} system  - that is when $\phi_{\nu,o}=\lim_{L_2\rightarrow L_1}\lim_{L_1 \rightarrow\infty} \phi_{\nu,o}(n\propto L_1)$ -  the $C_6$ symmetry of the superconducting gap is restored separately for the singlet and triplet channels. However, as we have stated recently \cite{Zegrodnik2024}, the coexistence of both the singlet ($d\pm id$) and triplet ($p\mp ip$) pairings results eventually in the overall gap characterized by $C_3$ symmetry, that is, the same as Hamiltonian with spin-valley locking considered here. In this view, the formation  of mixed $d_{xy}$/$p_y$ order in the narrow cylinder can be regarded as \emph{precursor} of the formation of $C_3$ symmetric superconducting gap in the bulk limit, as the outcome from calculations performed for  $L_1\times L_2=24\times4$ suggests. However, we believe that a systematic study in this respect should be performed in the future. This task is  demanding mainly due to the memory limitations to achieve it in the framework of DMRG method in its MPS formulation. Note that although we have not been able to obtain satisfactory accuracy affecting convergence in pairing correlation functions for the cylinder of size $48 \times 4$ even for $M=16000$, we observe \emph{tendency} toward the formation of $d\pm id$/$p \mp p$ superconducting orders in the considered case as well. Explicitly, for $n=20$ we find in the singlet channel: $\phi_s^2\approx 4 \times 10^{-6}$, $\phi_{d_{xy}}^2\approx 7\times10^{-5}$, $\phi_{dx^2-y^2}^2\approx 4\times 10^{-5}$, $\phi_{d+id:}^{2}\approx 1.5 \times 10^{-4}$, and $\phi_{d-id}^{2}\approx1.9\times 10^{-5}$; whereas in the triplet channel: $\phi_{p_x}^2\approx 2.4\times 10^{-5}$, $\phi_{p_y}^2 \approx 1.2 \times 10^{-6}$, $\phi_{p+ip}^{2} \approx 9 \times 10^{-6}$, $\phi_{p-ip}^2 \approx 4.5 \times 10^{-5}$, and  $\phi_f^{2}\approx  1.4 \times 10^{-5}$. This indicates again  that the  role of the cylinder width is important  (if not crucial) with respect to the character of pairing, and that $d\pm id$/$p\mp ip$ symmetries possibly dominate in the \emph{bulk}.

\begin{figure}
    \centering
    \includegraphics[width=0.75\linewidth]{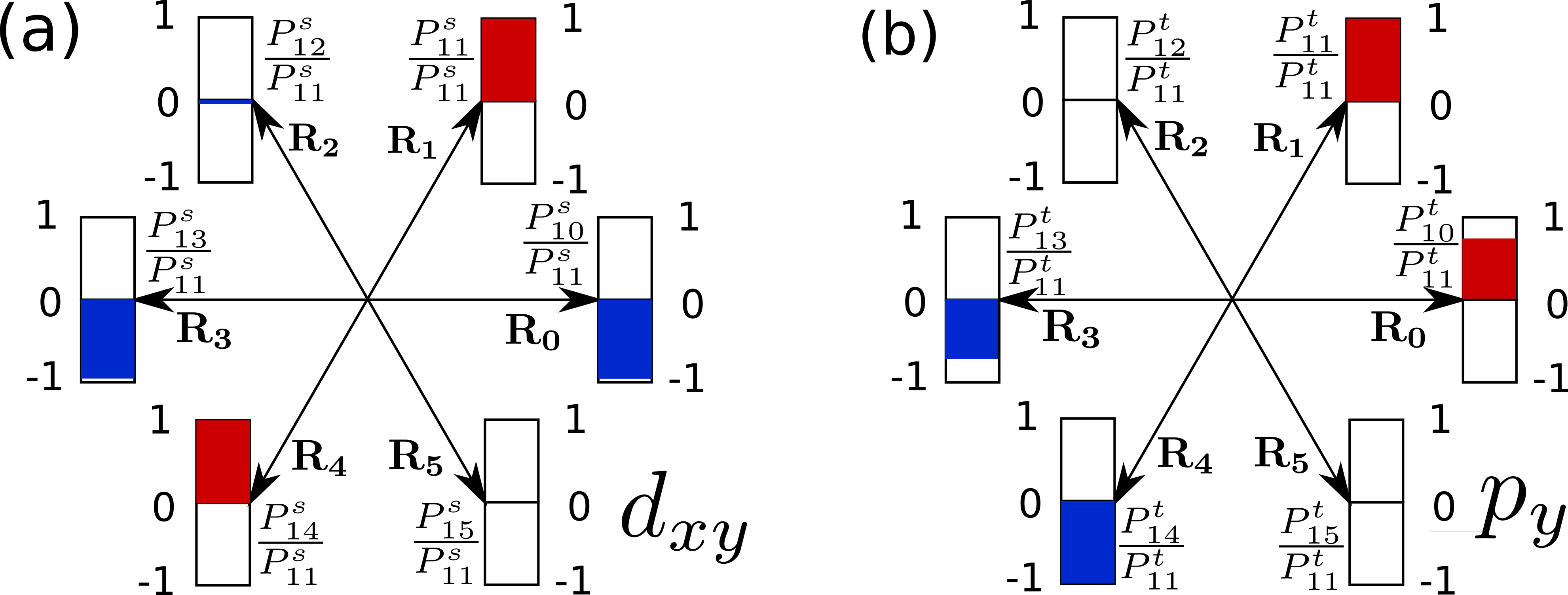}
     \caption{ The ratios of type $\frac{P_{\nu}^{1\beta}}{P_{\nu}^{11}}$ resulting from calculations carried out for the  cylinder $L_1\times L_2 = 48 \times3$, $D=0.4$ V/nm and $\delta\approx0.96$  for a singlet (a) and triplet (b) channels.
    }
    \label{fig:formfactors}
\end{figure}

\begin{figure}
    \centering
    \includegraphics[width=0.9\linewidth]{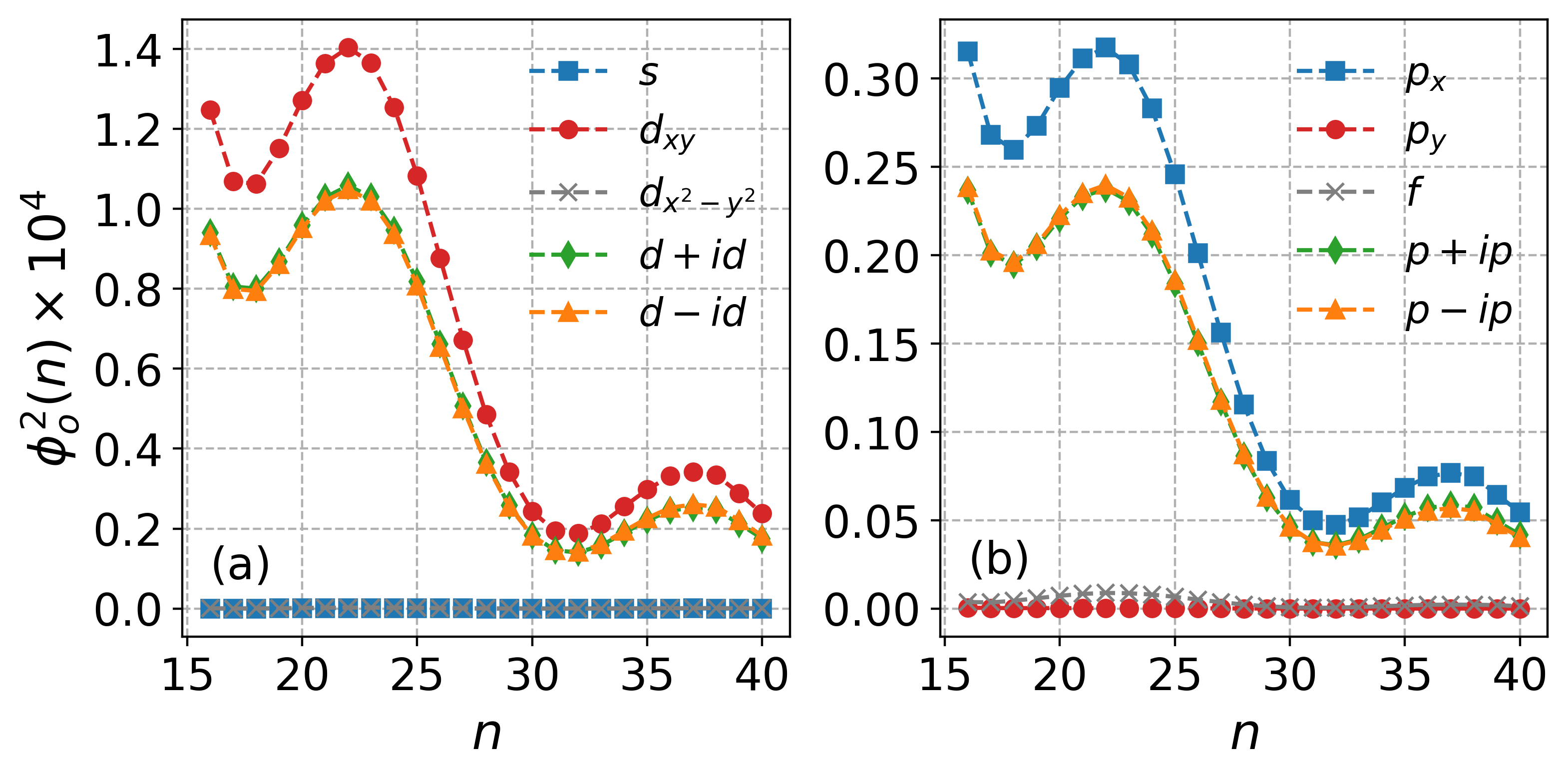}
     \caption{ The real-space resolved dependency of superconducting order parameters $\phi_o$ as a function of distance $n\mathbf{R_0}$  resulting from calculations carried out for the  cylinder $L_1\times L_2 = 48 \times3$, $D=0.4$ V/nm and $\delta\approx0.96$  for a singlet (a) and triplet (b) channels.
    }
    \label{fig:scorder48}
\end{figure}

\begin{figure}
    \centering
    \includegraphics[width=0.85\linewidth]{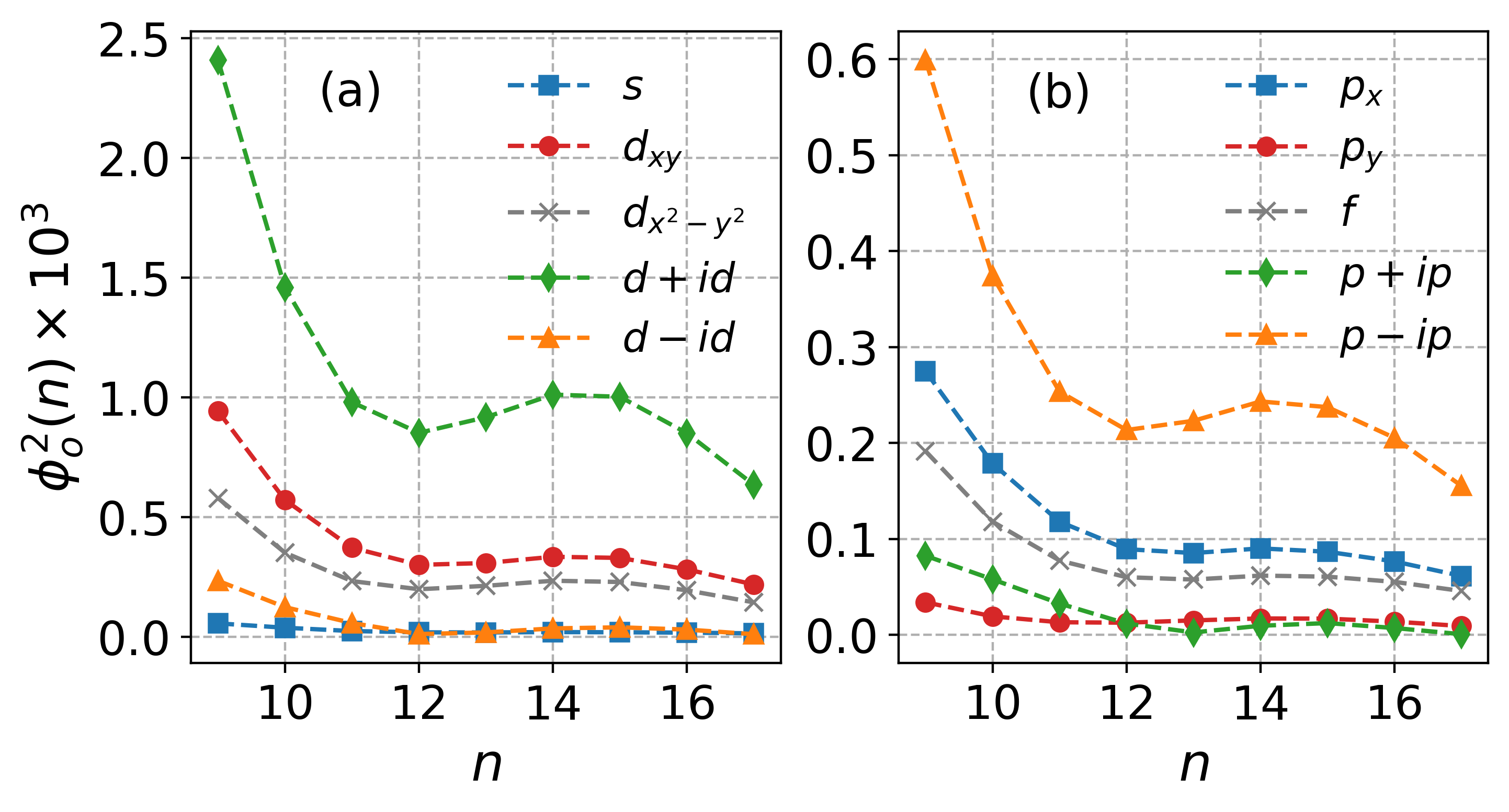}
     \caption{ The real-space resolved dependency of superconducting order parameters $\phi_o$ as a function of distance $n\mathbf{R_0}$  resulting from calculations carried out for the  cylinder $L_1\times L_2 = 24 \times4$, $D=0.4$ V/nm and $\delta\approx0.96$  for a singlet (a) and triplet (b) channels.
    }
    \label{fig:scorder24}
\end{figure}

\begin{figure}
    \centering
    \includegraphics[width=0.95\linewidth]{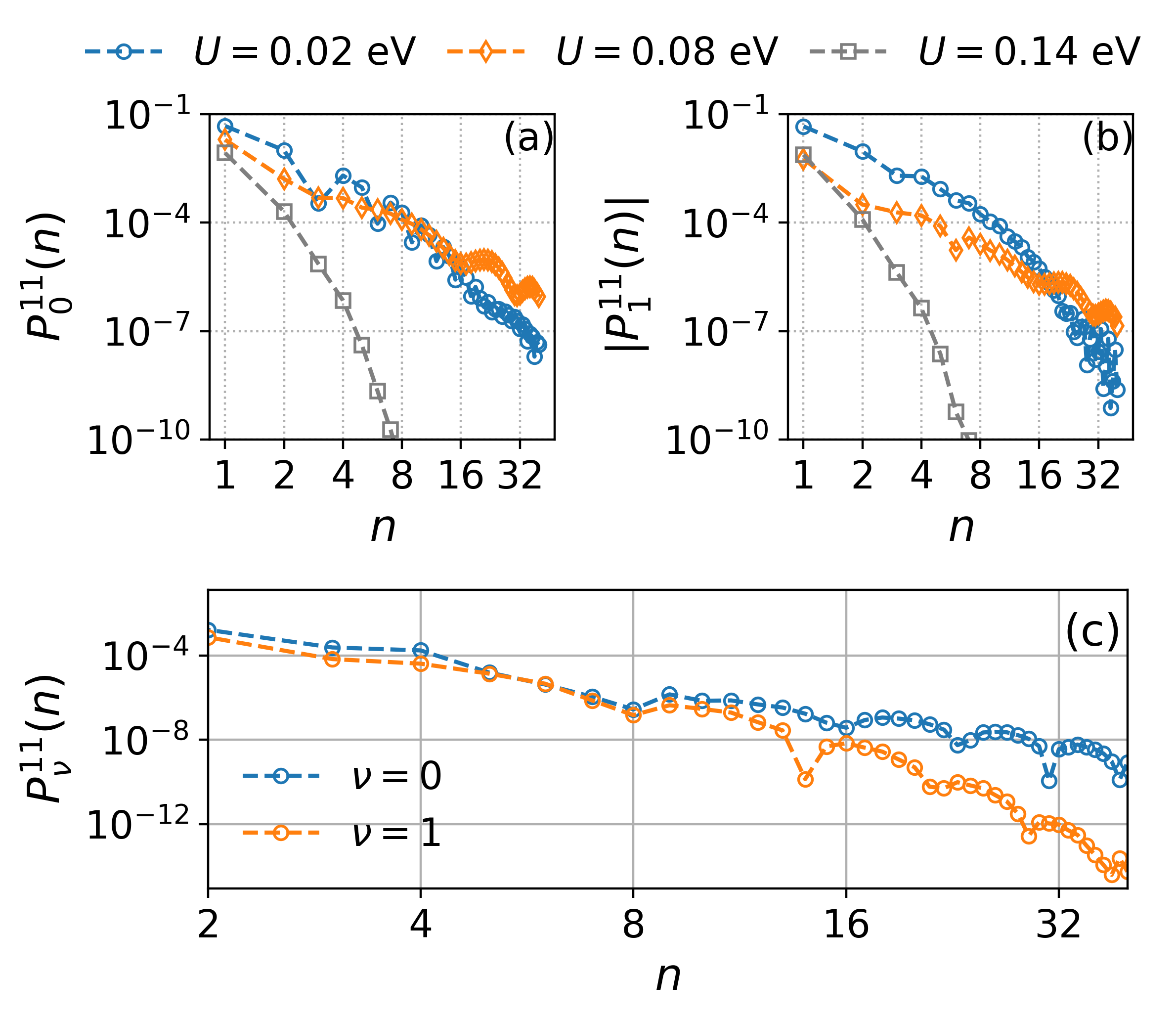}
     \caption{ The pairing correlation functions for singlet (a) and triplet channels (b) obtained for the three different values of $U$ referring to weakly-, moderately- and strongly correlated regimes for $\delta\approx0.96$ and $D=0.4$ V/nm. In (c) we show pairing correlation functions for the singlet and triplet channel obtained for $\approx0.96$ and $U=0.08$ eV, but the hopping amplitude is real and assumed to be negative of the absolute value of $t_{ij}^{D,\sigma}$ when $D=0.4$ V/nm.
    }
    \label{fig:Udep}
\end{figure}

Eventually we have qualitatively investigated the behavior of pairing correlation functions with respect to the value of $U$. That is, we have inspected their decay for $D=0.4$ V/nm and $\delta\approx0.96$, taking into account two opposite scenarios: when the system is weakly coupled by taking $U=0.02$ eV ($U<W$); and strongly correlated by setting $U=0.14$ eV ($U>W$). In Fig.~\ref{fig:Udep} we present the resulting pairings for both singlet (Fig.~\ref{fig:Udep}a) and triplet channels (Fig.~\ref{fig:Udep}b). Also, in Fig.~\ref{fig:Udep}c we present results obtained for $U=0.08$ eV   but for $t_{ij}^{D,\uparrow}=t_{ij}^{D,\downarrow}=-|t_{ij}^{D,\sigma}|$ that is in a scenario where effective spin-orbit coupling is absent in the Hamiltonian thus no spin-valley locking is present. Note that as for the triplet case  we obtain both positive and negative values, we show the absolute value of $P^{11}_{1}(n)$. We find that in the strongly correlated regime the superconducting correlations are suppressed and evidently decay exponentially. The exponential decay also appears in the triplet channel when $U=0.02$ eV. The magnitude of pairing correlation  functions in the singlet channel for the weakly correlated case is also substantially smaller when $n$ increases than that observed in the moderately correlated situation ($U=0.08\text{ eV} \approx W$). However, we were not able to state ultimately if the decay is also  of exponential character in this case. Anyway, our findings  suggest that enhancement in pairing is likely to take place for the moderately correlated scenario and is possibly caused by the delicate balance between interactions and the location of the van Hove singularity in $(\delta,D)$ space \cite{Zegrodnik2024,ZegrodnikBiborski}. Namely, we  detect weaker superconducting correlations in the singlet channel and exponential decay in triplet channel when spin-valley locking is absent (see Fig.~\ref{fig:Udep}c); in this case however, van Hove singularity  (vHS)refers to $\delta\approx1.5$ \cite{Zegrodnik2024} whereas for the spin-valley polarized case at $D=0.4$ V/nm vHS is located at $\delta\approx0.85$ \cite{Zegrodnik2024}, that is, close to the  superconducting state identified by us in $(\sigma,D)$ space. Therefore, both the enhanced density of states,  and sufficiently large electron-electron interactions possibly render the system to the superconducting state in WSe$_2$/WSe$_2$ homobilayer via coexisting singlet-triplet pairings harmonized with spin-valley polarization.

\emph{Summary.} In this Letter we provide indicators of the formation of the superconducting state in the effective spin-valley polarized Hubbard model describing the WSe$_2$/WSe$_2$ homobilayer. Both the decays of correlation functions and the relation between them indicate that the resulting superconducting state is of mixed singlet-triplet character and appears close to that reported in experiments on the ($\delta$,$D$)-plane. Our results possibly vindicate the role of electron-electron correlations in twisted materials in view of their superconducting properties. On the other hand, they provide the input into the evidence of ground-state properties of the spin-valley polarized single-band Hubbard Hamiltonian on the triangular lattice.


\begin{acknowledgments}
This research was funded by National Science Centre, Poland (NCN) according to decision 2021/42/E/ST3/00128. \\

We gratefully acknowledge Poland's high-performance Infrastructure PLGrid ACK Cyfronet AGH for providing computer facilities and support within computational grant no. PLG/2025/018169.

For the purpose of Open Access, the author has applied a
CC-BY public copyright license to any Author Accepted
Manuscript (AAM) version arising from this submission.

\end{acknowledgments}

\section{Supplementary Material}
\beginsupplement
\section{ Impact of bond dimension $M$ and the choice of the reference pair on the pairing correlation functions}
\begin{figure}
    \centering
    \includegraphics[width=0.95\linewidth]{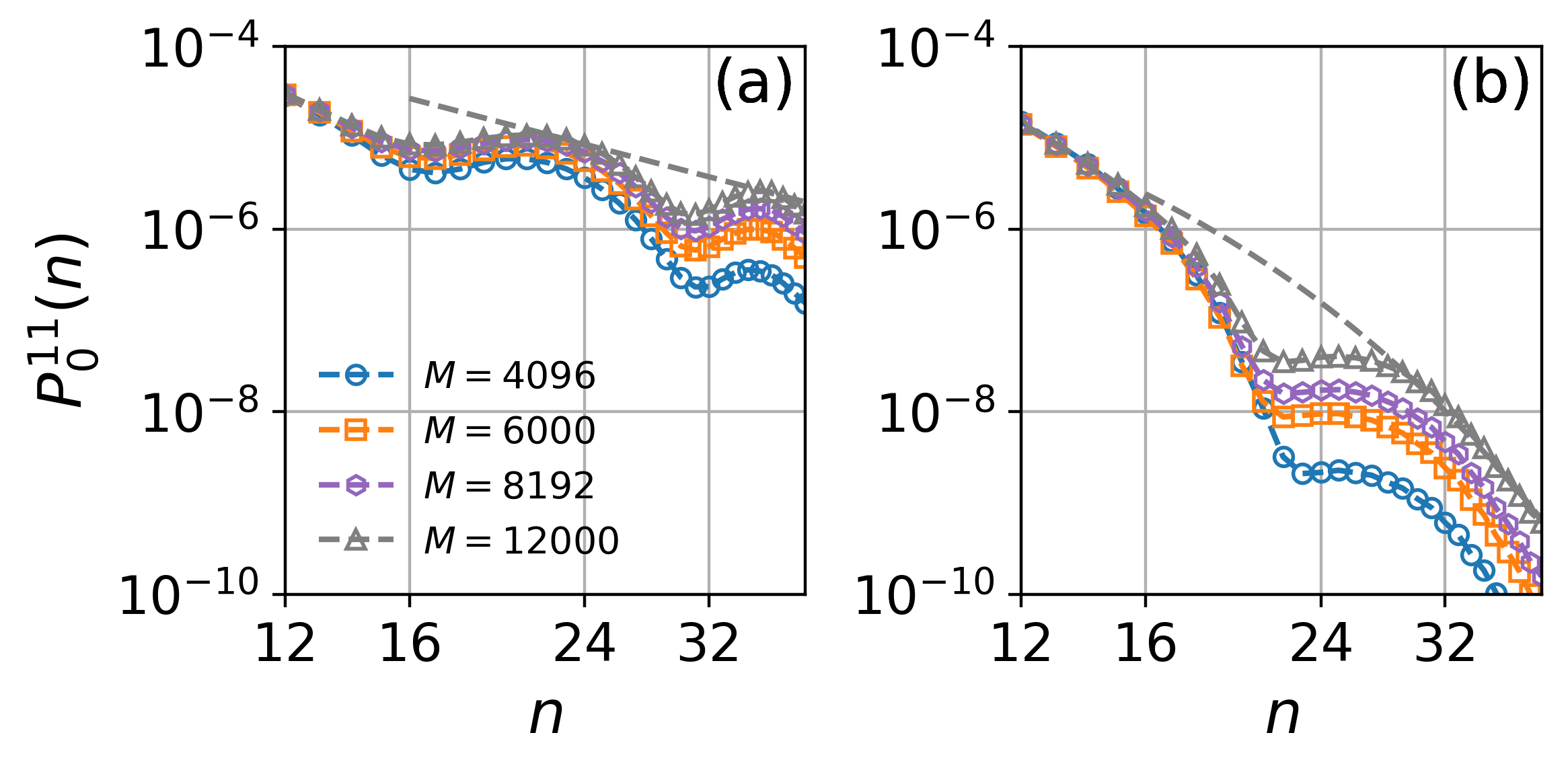}
     \caption{ The pairing correlation functions for singlet channel for $\delta\approx0.96$ for the different maximal bond dimension $M$. In (a) the displacement field is $D=0.4$ V/nm (power-law decay) whereas data for $D=0.5$ V/nm (exponential decay)  is presented in (b). Dashed gray lines represent $\sim n^{-K_s}$ (a) and $\sim\text{e}^{-n/\xi_s}$ (b) fits computed for $M=12000$.
    }
    \label{fig:A1}
\end{figure}

\begin{figure}
    \centering
    \includegraphics[width=0.95\linewidth]{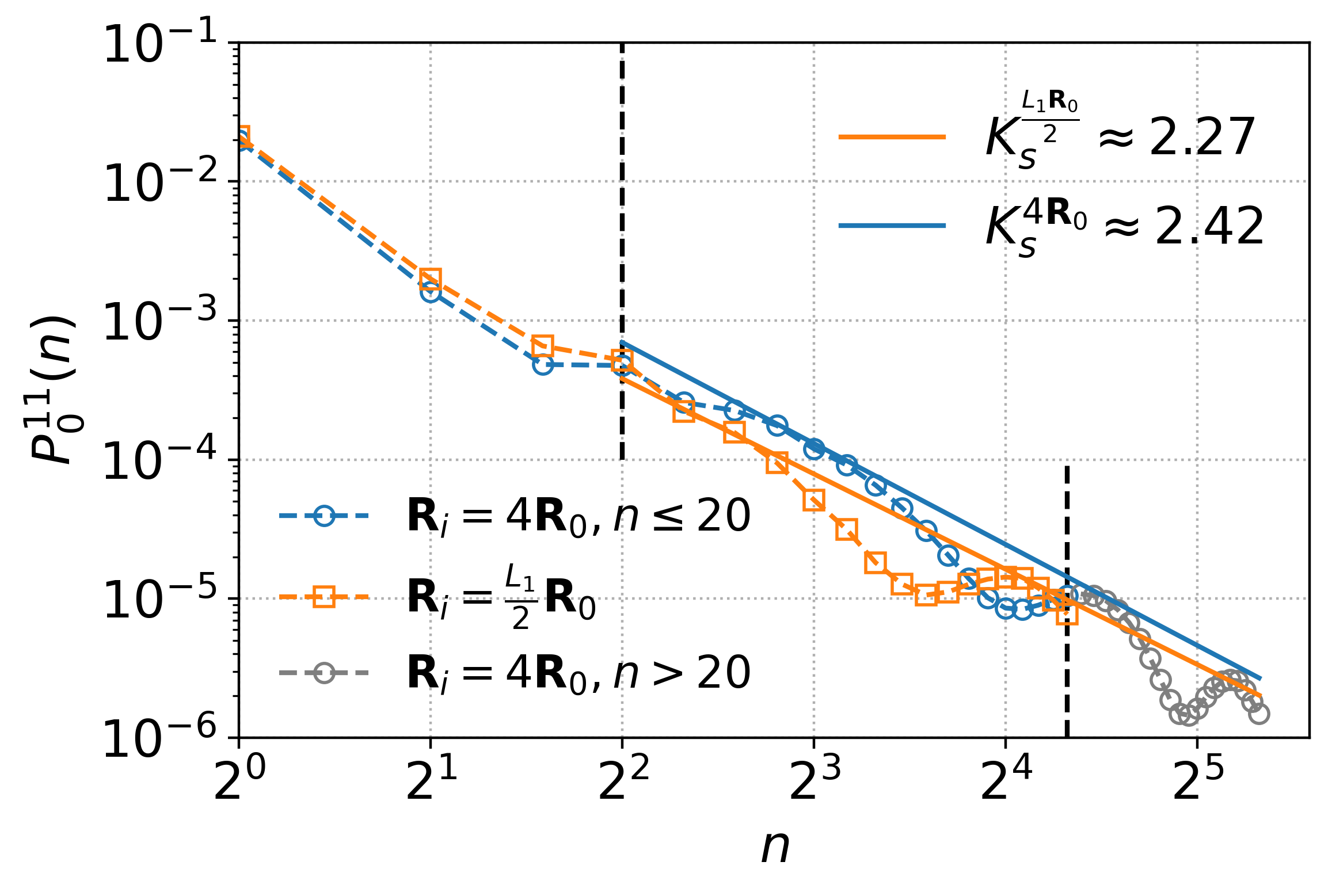}
     \caption{ The pairing correlation functions for $\delta\approx0.96$ and $D=0.4$ V/nm obtained for the two different selections of the reference pair index $i$ for the cylinder of size $L_1 \times L_2 = 48 \times 3$. Two vertical dashed lines indicate the range of $n$ for which fits have been performed, resulting in the similar values of exponents $K_s$ for the different choices of the reference pair index $i$. Solid color lines refer to fits $P^{11}_{0}(r=n|\mathbf{R}_0|)\sim r^{-K_s}$.
    }
    \label{fig:A2}
\end{figure}

The analysis regarding pairing correlation functions $P_{\nu}^{\alpha\beta}(n)\equiv\langle\hat{\Delta}^{\dagger}_{\nu,\alpha}(\mathbf{R}_i)\hat{\Delta}_{\nu,\beta}(\mathbf{R}_i+n\mathbf{R}_0)\rangle$ is crucial in view of identification if quasi-long range superconducting order emerges for given $(\delta,D)$. As the charge  density profiles along the cylinder are oscillating, the $P^{\alpha\beta}_{\nu}(n)$ does so too. Therefore one may consider if the choice of index $i$ labeling the reference pair affects the interpretation. Undoubtedly it may happen when $i$ is to close to the cylinder edge and in turn correlation functions are affected strongly by open boundaries of the cylinder. On the other hand, we intended to investigate the pairing for the largest reliable $n$. To achieve this we have set $i$ referring to the site located at $4\mathbf{R}_0$ and computed $P_{\nu}^{\alpha\beta}(n)$ up to $n=40$, that is for the site at position $L_1\mathbf{R}_0-4\mathbf{R}_0$. Subsequently we have performed calculations with a maximal bond dimension in MPS  being $M=4096,6000,8192$ and $12000$. This provided us with the evidence of the enhancement of the amplitudes of long-ranged pairing saturating with increasing value of $M$ as shown in Fig.~\ref{fig:A1}. Then we have set $i$ to be close to the center of the length ($L_1)$ of the cylinder, that is $\mathbf{R}_i=\frac{L_1}{2}\mathbf{R}_0$ and computed $P^{11}_0(n)$ up to $n=L_1\mathbf{R}_0-4\mathbf{R}_0$. As we performed this analysis for $D=0.4$ V/nm and $\delta\approx0.96$, this choice of $i$ refers to a minimum in charge density (cf. Fig.~6a in the main text). Then,  we have compared $P^{11}_0(n)$ for both cases as shown in Fig.~\ref{fig:A2}. Disregarding phase shift the obtained pairing correlation functions are close to each other. Specifically, we have performed $P^{11}_0(n)\propto r^{-K_s}=(n|\mathbf{R}_0|)^{-K_s}$ fits for $n\in(8,20)$ which resulted in $K_s\approx2.27$ when $i\sim L_1/2$ and $K_s\approx2.42$ when $i\sim4|\mathbf{R}_0|$. Therefore we conclude that the selection of the location of the reference site for computing correlation functions only slightly modifies the estimated asymptotic behavior which is not qualitatively sensitive to the charge modulations and presence of the open boundaries in view of the analysis applied by us.

\section{The error from fitted values}
\begin{figure}
    \centering
    \includegraphics[width=0.95\linewidth]{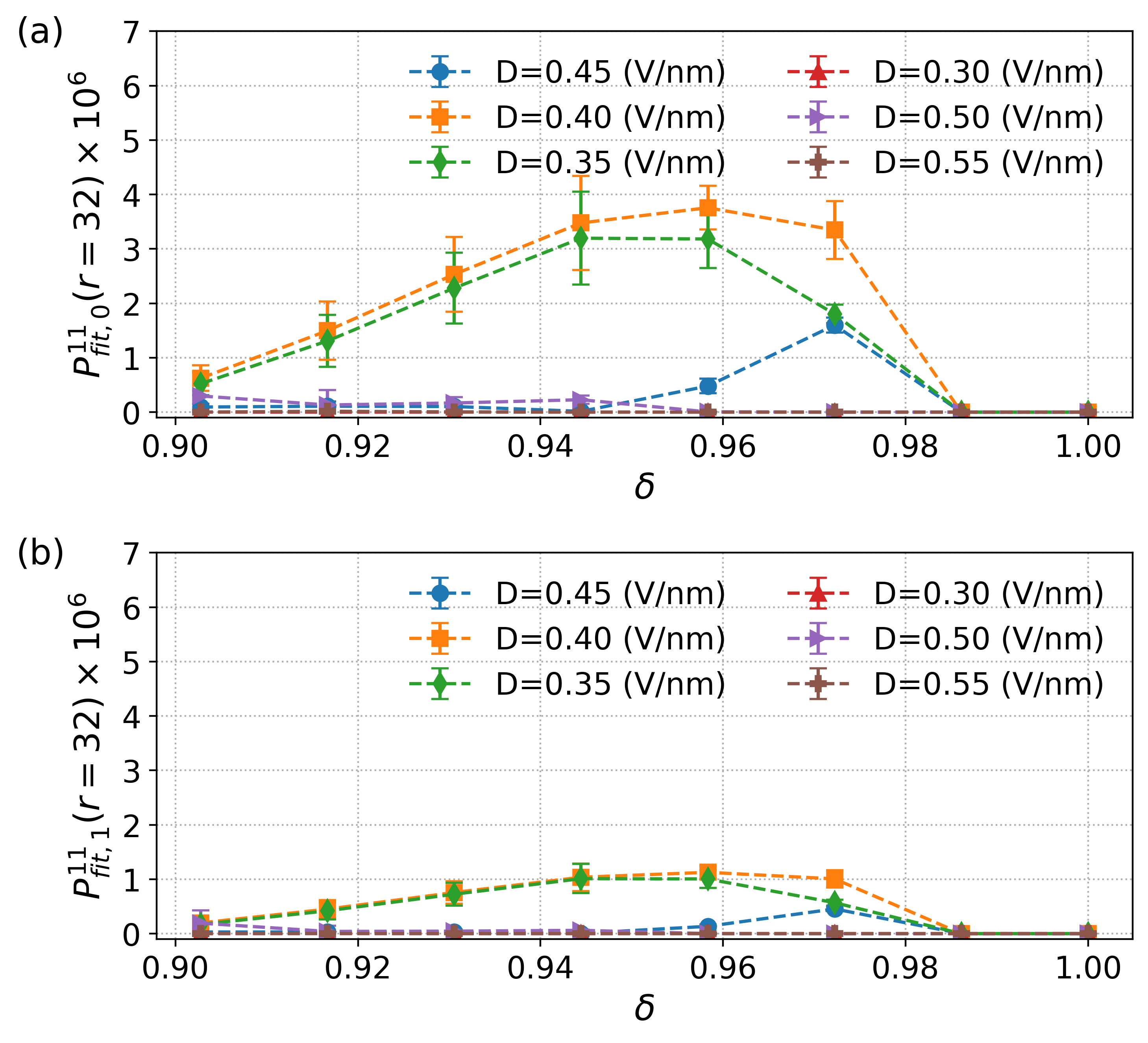}
     \caption{ The pairing correlation functions $\text{max}[P^{11}_{fit,\nu}]$ for $n=32$ and corresponding errors resulting from fitting procedure both for singlet (a) and triplet (b) channels obtained for the considered values of hole doping and displacement field.
    }
    \label{fig:A3}
\end{figure}

In Fig.~4 in the main text, we present the phase diagram in $(\delta,D)$ space based on $\text{max}[P^{11}_{fit,\nu}]$ obtained from fits. As the fitting procedure provides errors (e.g. in $K_s$ exponent) they propagate into the values of $P^{11}_{fit,\nu}$. However, we find that those errors are relatively small and therefore do not invalidate the phase diagram as shown in Fig.~\ref{fig:A3}. Furthermore, when considering the minimal value of the signal, that is $\text{min}[P^{11}_{fit,\nu}]$, we still obtain the same landscape in $(\delta,D)$.
\section{Identification of pairing order}
The identification of pairing order can be achieved on the basis of the real-space resolved  superconducting order parameter $\phi_{\nu,o}=\sqrt{\lim_{n\rightarrow\infty}\phi_{\nu,o}^{2}(n)}$,  where
\begin{equation*}
    \begin{split}
    \phi_{o}^{2}(n)=&\Bigg|\Big\langle\sum_{\alpha,\beta}f_{o}^{\alpha}f_{o}^{\beta}\hat{\Delta}_{\nu(o)\alpha}^{\dagger}(\mathbf{r_i})\hat{\Delta}_{\nu(o)\beta}(\mathbf{r_i}+n\mathbf{R_0})\Big\rangle\Bigg|\\
    =&\Bigg|\sum_{\alpha,\beta}f_{o}^{\alpha}f_{o}^{\beta}P_{\nu(o)}^{\alpha,\beta}(n)\Bigg|,
    \end{split}
\end{equation*}
$f_{\nu,o}^{\alpha}$ are \emph{form-factors} which allow  to identify the order by inspecting the maximum of $\phi_{\nu,o}(n)^2$ for a given set $P_{\nu}^{\alpha,\beta}$ resulting from calculations with respect to the possible order provided by index $o$. The possible pairings are $o\in\{s,d_{xy},d_{x^2-y^s}\}\cup\{d_{x^2-y^2} \pm id_{xy}\}$  in the singlet channel, and $o\in\{p_x,p_y,f\} \cup \{p_y\pm ip_{x}\}$ in the triplet channel. Note that since we work in a regime assuming a constant number of particles for a given $\delta$ we \emph{need} to inspect the square of $\phi_{\nu,o}(n)$ as it is the expectation value of the operator conserving the number of electrons considered. Therefore, we have a gauge in the global sign of form factors. Also, there is a gauge in relating the form factors to the rotation of the frame of reference which influences their magnitude; however, as we find for cylinder $L_1\times L_2=48\times 3$ that pairing correlation functions in which $\alpha,\beta\in\{2,5\}$ are substantially smaller than the remaining (about two orders of magnitude), we are still able to identify the pattern of pairing order assuming $f_{\nu,o}^{2}=f_{\nu,o}^{5}=0$. Eventually for longer and narrower cylinders we inspect the $\phi_{o}^2(n)$ for a rotation gauge $k=0$ according to the convention of $\{f_{o}^{\alpha}\}$ provided in Tab.\ref{tab:formafactors}. This selection affects the amplitudes of  $\{d_{xy},d_{x^2-y^s}\}$ and $\{p_x,p_y\}$ pairings; however, it does not $s$ and $f$ order parameters. The order parameters of $d\pm id$ and $p \pm ip$ symmetries are provided by form factors given by
\begin{align*}
    f_{o}^{\alpha}=i^{c_o}e^{im_{o}\varphi_{\alpha}},
\end{align*}
where $\varphi_\alpha=\angle(\mathbf{R}_{\alpha},\mathbf{R}_0)\in\{0,\pi/3,2\pi/3,\pi,4\pi/3,5\pi/3\}$, and $c$,$m$ are listed in Tab. \ref{tab:formafactorsC}. 

\begin{table}[h]
    \centering
    \begin{tabular}{c | c | c | c | c | c | c  }
        
         $f^{\alpha}_{o}$& $\mathbf{R}_{l_k}$ & $\mathbf{R}_{l_{k+1}}$ & $\mathbf{R}_{l_{k+2}}$ & $\mathbf{R}_{l_{k+3}}$ & $\mathbf{R}_{l_{k+4}}$ & $\mathbf{R}_{l_{k+5}}$ \\
        \hline \hline
        $s$  & 1 & 1 & 1 & 1 & 1 & 1 \\  \hline
        $d_{xy}$& 1 & -1 & 0 & 1 & -1 & 0  \\ \hline
        $d_{x^2-y^2}$& -1 & -1 & 2 & -1 & -1 & 2  \\ \hline \hline
        $f$  & -1 & 1 & -1 & 1 & -1 & 1 \\  \hline
        $p_{y}$& -1 &- 1 & 0 & 1 & 1 & 0  \\ \hline
        $p_{x}$& -1 & 1 & 2 & 1 & -1 & -2  \\ \hline \hline

    \end{tabular}
    \caption{Form factors of superconducting order for singlet and triplet channels utilized in the presented analysis. Note that $l_m=m \mod{6}$.}
    \label{tab:formafactors}
\end{table}

\begin{table}[h]
    \centering
    \begin{tabular}{c | c | c || c | c  }
        
         $f^{\alpha}_{o}$& $d+id$ & $d-id$ & $p+ip$ & $p-ip$ \\
        \hline \hline
        $c$  & 0 & 0 & 1 & 1 \\  \hline
        $m$  & 2 & 4 & 1 & 5 \\ \hline \hline
              
    \end{tabular}
    \caption{Parameters $c_o$ and $m_o$ in form factors $f_o^{\alpha}$ describing superconducting order of $d\pm id$ and $p\pm ip$ symmetries.}
    \label{tab:formafactorsC}
\end{table}

\bibliography{refs.bib}

\end{document}